\documentclass{article}
\usepackage{epsf}
\usepackage{amssymb}
\usepackage[T2A]{fontenc}
\usepackage[english]{babel}
\usepackage[dvips]{graphicx}

\begin{document}
\title{De Sitter stability in quadratic gravity}
\author{A. V. Toporensky$^{\dagger}$ and P. V. Tretyakov$^{\ddagger}$}
\date{}
\maketitle
\hspace{8mm}{\em Sternberg Astronomical Institute,
Universitetsky prospect, 13,  Moscow 119992, Russia}

\bigskip

$^{\ddagger}$ Electronic mail: tpv@xray.sai.msu.ru\\
$^{\dagger}$ Electronic mail: lesha@sai.msu.ru\\

\begin{abstract}
Quadratic curvature corrections to Einstein-Hilbert action lead
in general to higher-order equations of motion, which can induced
instability of some unperturbed solutions of General Relativity.
We study conditions for stability of de Sitter cosmological 
solution. We argue that simple form of this condition known for 
FRW background in $3+1$ dimensions changes seriously if at least
one of these two assumptions is violated. In the present paper 
the stability conditions for de Sitter solution have been found
for multidimensional FRW background and for Bianchi I metrics in 
$3+1$ dimensions.
\end{abstract}

\section{Introduction}
Recently theories of higher-order gravity have become a matter
of intensive investigations. Some of them are motivated by string theory
 \cite{Tseytlin},
other proposals have been put forward in order to explain present 
accelerated expansion of the Universe (see \cite{Odintsov,Odintsov2}
and references therein). One of the main features of these
theories is the fact that they generally lead to equations of motion which 
contain higher-order derivatives.
In $3+1$ dimensions all higher-curvature terms in Lagrangian
which keep the equations of motion to be the same order differential equations
as in General Relativity vanish \cite{Lovelock}. In string gravity the second order correction
contains dilaton scalar field in addition to the gravity sector. These
corrections  
has a particular structure
(a product of Gauss -- Bonnet term and a function of a dilaton field)
which prevents the resulting equations of motion from being of higher-order
equation.
However, these equations become
higher-derivative starting from the 3-d order corrections (see,
for example, \cite{Bento}.)

As the number of time derivatives in the equations of motions increases in comparison
with the situation in classical General relativity, the dimensionality of the phase
space of the corresponding dynamical system increases as well. This may result in 
a dynamical behavior qualitatively different from that known in GR: some
important GR solutions
can loose their stability (this effect from a general point of view have
been described in \cite{Smilga}). 
Though higher-derivative terms may be small on classical solution,
perturbations from higher-order corrections 
can grow in direction of "additional" dimensions of the phase space,
and a point which is stable in GR may become unstable in higher-order theory. 
This situation has
already been described in 4-th order string gravity for Schwarzschild \cite{Alexeyev}
and de Sitter solutions  \cite{Sami}, as well as in cosmological scenarios with
vacuum polarization \cite{we}. The other proposal of this type is so called
modified gravity with corrections in the form of function of the curvature
$f(R)$ \cite{Carloni, Carroll} which is also not free from instabilities
\cite{Dolgov, Faraoni} (see, however, \cite{Odintsov3} and recent rewiews \cite{
Odintsov4, Odintsov5}).

Recent understanding that modern string theories may admit many
(actually, {\em very} many) solutions in which 4-D part is the
(metastable) de Sitter space-time with a positive effective
cosmological constant (\cite{KKLT1, KKLT2} and many following papers) renew
interest in the investigation of the stability of de Sitter
solutions in gravity theories containing higher order powers
of the Riemann tensor in their effective Lagrangian. These terms
were long known to arise from quantum-gravitational corrections
to the Einstein gravity (either the vacuum one, or interacting with matter
quantum fields). Presumably, some of them may remain in the string
theory in spite of many cancellations due to its numerous symmetries.
Many particular forms of second order curvature terms have been
studied in connection to possible instabilities induced by them
\cite{Star, Muller, Schmidt}. This work have 
been also extended to $R^n$ gravity \cite{Dunsby1, Dunsby2}.
In the present paper we study stability of de Sitter solution
with respect to second order curvature corrections in their general form.

\section{FRW Universe in $3+1$ dimension}
A general form of second order curvature corrections is

\begin{equation}
S_2=\int\sqrt{-\mathrm{g}} (\alpha R^{iklm}R_{iklm}+\beta
R^{ik}R_{ik}+\gamma R^2)\,d^4x.
 \label{1}
\end{equation}

However, in $3+1$ dimensions the Gauss-Bonnet term
$$GB= R^{iklm}R_{iklm}-4R^{ik}R_{ik}+R^2$$
does not contribute to equations of motion, and, then,
we can rewrite (1) in the form

\begin{equation}
S_2=\int \sqrt{-\mathrm{g}} (\alpha R^{iklm}R_{iklm}+\beta
R^{ik}R_{ik}+\gamma R^2-\alpha GB)\,d^4x.
 \label{2}
\end{equation}

Introducing 
$B=\beta+4\alpha$, $C=\gamma-\alpha$ we get the following action for
the theory with a cosmological constant term
\begin{equation}
S=\int \sqrt{-\mathrm{g}} (R +B R^{ik}R_{ik}+C R^2-
\Lambda)\,d^4x.
 \label{3}
\end{equation}

For the Friedmann-Robertson-Walker metrics
\begin{equation}
\mathrm{g}_{ik}=diag(-n(t)^2,a(t)^2,a(t)^2,a(t)^2).
 \label{4}
\end{equation}
we have
$$
R=\frac{6}{n^2}\left[\frac{\ddot a}{a}-\frac{\dot a}{a}\frac{\dot
n}{n}+\frac{\dot a^2}{a^2} \right],
$$
$$
R_{ik}^2=\frac{12}{n^4}\left[\frac{\ddot a^2}{a^2}-\frac{\ddot
a}{a}\frac{\dot a}{a}\frac{\dot n}{n}+\frac{\dot
n^2}{n^2}\frac{\dot a^2}{a^2}+\frac{\ddot a}{a}\frac{\dot
a^2}{a^2}-\frac{\dot a^3}{a^3}\frac{\dot n}{n}+\frac{\dot
a^4}{a^4} \right].
$$

Varying the action (3) with respect to $n$ and setting $n=1$ we
obtain the following analog of Friedmann equation
(we denote
$D=12B+36C$):
\begin{equation}
6H^2-\Lambda +2DH^2(\dot H+H^2)-3DH^4-D(\dot H+H^2)^2+2DH(\ddot
H+3H\dot H+H^3)=0,
 \label{5}
\end{equation}
where $H\equiv\frac{\dot a(t)}{a(t)}$ is the Hubble parameter.

Note that the Friedmann equations, being algebraic in GR, becomes
a differential one in the theory with second order curvature
corrections  when $D \ne 0$. In the $D=0$ case the Friedmann equation
does not change at all, and we exclude this case from the further analysis.

For stability studies this equation should be written in the form
of a system of two first order equations:

\begin{equation}
\begin{array}{l}
\dot H =F,\\

\dot F = -3HF-H^3+\frac{1}{2DH}(\Lambda - 6H^2+2DH^4+DF^2) \equiv f.

\end{array}
 \label{6}
\end{equation}

The form of de Sitter solution ($\dot H =0$,$\dot F =0$) give us the equations
for corresponding stable points:
$$
-H^3_0+\frac{1}{2DH_0}(\Lambda - 6H_0^2+2DH_0^4)=0,
$$
its solution is
$H_0=\pm\sqrt{\frac{\Lambda}{6}}$.
It is remarkable, that this de Sitter solution is exactly the same
as in the pure GR, so higher-order corrections in the theory under
investigation do not shift the location of the fixed point. However,
its stability can be affected by the higher-order terms.

Linearizing (\ref{6}) we obtain

\begin{equation}
\begin{array}{l}
\dot H =F,\\

\dot F = \left ( \frac{\partial f}{\partial F}\right)_0F+\left (
\frac{\partial f}{\partial H}\right)_0H,

\end{array}
 \label{7}
\end{equation}

where
$$
\frac{\partial f}{\partial F}=-3H+\frac{F}{H},
$$
$$
\frac{\partial f}{\partial
H}=-3F-3H^2-\frac{\Lambda-6H^2+2DH^4+DF^2}{2DH^2}+\frac{-12H+8DH^3}{2DH}.
$$

Two eigenvalues of this system are
$$
 \mu_{1,2}=\frac{1}{2}[(\frac{\partial f}{\partial F})_0\pm \sqrt{(\frac{\partial f}{\partial F})_0^2 +4(\frac{\partial
f}{\partial H})_0} ].
$$

For stability of the solution it is necessary that both eigenvalues have
negative real parts. Substituting the de Sitter solution we get that
in a stable point $\left ( \frac{\partial f}{\partial F}\right)_0 = -3H_0$.

As
$\left ( \frac{\partial f}{\partial F}\right)_0<0$
in an expanding Universe, negativity of eigenvalues requires 
$\left ( \frac{\partial f}{\partial H}\right)_0<0$. 
On the other hand, we have from (7) that in the de Sitter point
$$\left (
\frac{\partial f}{\partial H}\right)_0=\frac{-6}{D},
$$
which means that the condition for stability of the de Sitter
solution is
$D= 12(\alpha+\beta+3\gamma)>0$.

In is possible to consider this result from a different point of view. Instead of
Riemann tensor, we can express second order corrections through Weil tensor
$$
C_{iklm}^2=R_{iklm}^2+(N-6)R_{ik}^2+\left(\frac{7}{3}-\frac{13}{18}N+\frac{1}{18}N^2
\right)R^2,
$$
where $N$ is the dimensionality of space-time
and the curvature scalar:

\begin{equation}
S=\int \sqrt{-\mathrm{g}} (R +\tilde{B} C^{iklm}C_{iklm}+\tilde{C}
R^2- \Lambda)\,d^4x,
\end{equation}
where $\tilde{B}=2B=2\alpha+\frac{1}{2}\beta$ and

$\tilde{C}=\frac{1}{3}(B+3C)=\frac{1}{3}(\alpha+\beta+3\gamma)=D/36$.

The Weil tensor vanishes on the Friedmann metrics,
and the general quadratic corrections reduce to the $R^2$-term
with the known result: 
the de Sitter solution
is stable if the coefficient before $R^2$-term in (8) is positive.

\section{Multidimensional FRW Universe}

It is instructive to compare this result with the situation in higher
dimensional theory, where the Gauss-Bonnet term does contribute 
to dynamical equations. To illustrate importance of this property,
we consider the theory with second order curvature corrections, which we write down in the
form

\begin{equation}
S=\int\sqrt{-\mathrm{g}} (R + A C^{iklm}C_{iklm}+B
R^{ik}R_{ik}+C R^2-\Lambda)\,d^Nx,
 \label{17}
\end{equation}

As usual, 
$C_{iklm}=0$ on the Friedmann metrics,
and we have two independent parameters of the theory $B$ and $C$.
The explicit forms of second order corrections on the Friedmann metrics are

\begin{equation}
R=\frac{(N-1)}{n^2}\left[2\frac{\ddot a}{a}-2\frac{\dot
n}{n}\frac{\dot a}{a}+(N-2)\frac{\dot a^2}{a^2}\right],
 \label{20}
\end{equation}

\begin{equation}
R_{ik}^2\!\!=\!\!\frac{(N-1)}{n^4}\!\!\left[\!N\frac{\ddot
a^2}{a^2}\!-\!\!2N\frac{\dot n}{n}\frac{\dot a}{a}\frac{\ddot
a}{a}\!+\!\!N\frac{\dot n^2}{n^2}\frac{\dot a^2}{a^2}
\!+\!\!2(N\!-\!2)\frac{\ddot a}{a}\frac{\dot
a^2}{a^2}\!-\!\!2(N\!-\!2)\frac{\dot n}{n}\frac{\dot
a^3}{a^3}\!+\!\!(N\!-\!2)^2\frac{\dot a^4}{a^4}\right],
 \label{21}
\end{equation}

\begin{equation}
R_{iklm}^2=\frac{2(N-1)}{n^4}\left[2\frac{\ddot
a^2}{a^2}-4\frac{\dot n}{n}\frac{\dot a}{a}\frac{\ddot
a}{a}+2\frac{\dot n^2}{n^2}\frac{\dot a^2}{a^2}+(N-2)\frac{\dot
a^4}{a^4}\right].
 \label{22}
\end{equation}

Varying (\ref{17}) with respect to $n$, setting $n=1$
and solving the equation of motion with respect to the second
derivative we get

\begin{equation}
\begin{array}{l}
\ddot H=\frac{-1}{2H(N-1)[BN+4C(N-1)]}[(N-1)(N-2)H^2 - \Lambda +
B\{(N-1)^2(N-4)H^4\\
\\
 -N(N-1)\dot H^2+2N(N-1)^2\dot HH^2\}+C\{N(N-1)^2(N-4)H^4-4(N-1)^2\dot H^2\\
 \\
 +8(N-1)^3\dot HH^2\}]\equiv f.
\end{array}
 \label{23}
\end{equation}

It should be noted that 
if $BN+4C(N-1)=0$ the resulting equation of motion does not contain $\dot H$ and
$\ddot H$ terms, which means that a new de Sitter solution is trivially stable. This case
should be excluded from our analysis.

The de Sitter fixed points can be found from the equation
\begin{equation}
(N-1)^2(N-4)(B+CN)H_0^4+(N-1)(N-2)H_0^2-\Lambda=0,
 \label{24}
\end{equation}

and have the form
\begin{equation}
H_0^2=\frac{-(N-2)\pm\sqrt{(N-2)^2+4\Lambda(N-4)(B+CN)}}{2(N-1)(N-4)(B+CN)}
 \label{25}
\end{equation}

Comparing these results with those obtained in Sec.2 we can see
two important differences from the previously studied case of $3+1$ dimensions:
\begin{itemize}
\item The de Sitter points are shifted in comparison with classical point 
$H^2=\Lambda/((N-1)(N-2))$.
It happens due to the first term in eq.(\ref{24}) which vanishes in $N=4$ dimensions.
\item There are {\it two} different de Sitter solutions, corresponding to two positive
roots of (\ref{25}). Each of them has its own zone of existence, so depending on $B$ and $C$
we can have two or one de sitter solutions, or the situation when de Sitter solutions are absent.

\end {itemize}
 The point with $(+)$ sign  in (\ref{25}) has a regular behavior for 
$B \to 0$, $C \to 0$ (it tends to unperturbed de Sitter solution),
the behavior of the point with $(-)$ sign is singular.
The latter point can not be an analog of any de Sitter point, existing
in the pure GR, and exists even in the case of $\Lambda=0$.

For stability analysis we should rewrite (\ref{23}) 
in the form of dynamical system (7). Using the same notations
as in the previous section, we get 
$$\left ( \frac{\partial f}{\partial F}\right)_0=-(N-1)H_0,$$
and, for the same reason, the condition for de Sitter stability is equivalent to negativity
of 
$\left ( \frac{\partial
f}{\partial H}\right)_0$.
The eq.(\ref{23}) gives
$$
\begin{array}{l}
\left ( \frac{\partial f}{\partial
H}\right)=-\frac{f}{H}-\frac{1}{2H(N-1)[BN+4C(N-1)]}[2(N-1)(N-2)H
+B\{4(N-1)^2(N-4)H^3\\
\\+4N(N-1)^2\dot H H\}+C\{4N(N-1)^2(N-4)H^3+16(N-1)^3\dot HH\}].
\end{array}
$$

In the de Sitter point we get
\begin{equation}
\left ( \frac{\partial f}{\partial
H}\right)_0=\mp\frac{\sqrt{(N-2)^2+4\Lambda(N-4)(B+CN)}}{BN+4C(N-1)},
 \label{26}
\end{equation}

We get that stability conditions depend on the sign of the combination $E=
BN+4C(N-1)$.
If it is positive,
the regular de Sitter point (if exists) is always stable, while
the singular de Sitter point is always unstable, independently of $\Lambda$.
If it is negative, the regular point is unstable, and the singular point is stable.

The conditions for existence of these de Sitter points can be easily derived
from (15). Two conditions which should be satisfied are: 
$(N-2)^2+4\Lambda(N-4)(B+CN)>0$
and $H_0^2>0$. Corresponding zones on $(B, C)$ plane are shown in Fig.1.
\begin{figure}[h]

\includegraphics[width=1\textwidth]{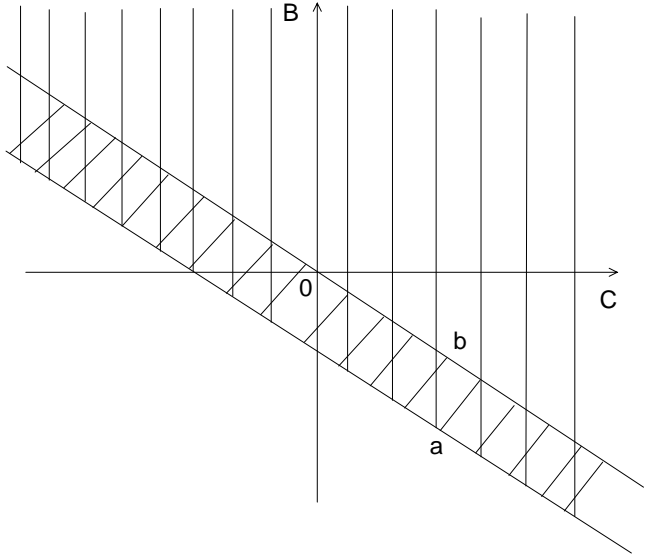}

   \label{f1}
\caption{Zones of de Sitter solutions for the action (9). A singular de Sitter exists
in double shaded band zone between line {\bf a} [$B+CN>-(N-2)^2/(4\Lambda
(N-4))$] and {\bf b} [$B+CN>0$], 
a regular de Sitter exists in both double shaded band and shaded
zone above the line {\bf b}.}
  \end{figure}

 Combining these
results with the stability analysis we can distinguish 
the following zones in the
parameter space $(B,C)$ (see Fig.2):
 \begin{itemize}
\item{\it for negative E}: 1) Zone with one unstable regular de Sitter ($NOR$),
 2) zone with one stable (singular) and 
one unstable (regular) de Sitter ($ROSK$), 3) zone without de Sitter solutions($KSM$).
\item{\it for positive E}: 1) Zone with one stable regular de Sitter ($PON$),
 2) zone with one stable (regular) and one unstable (singular)
de Sitter ($POSL$), 3) zone without de Sitter solutions ($LSM$).   
\end{itemize}

\begin{figure}[h]

\includegraphics[width=1\textwidth]{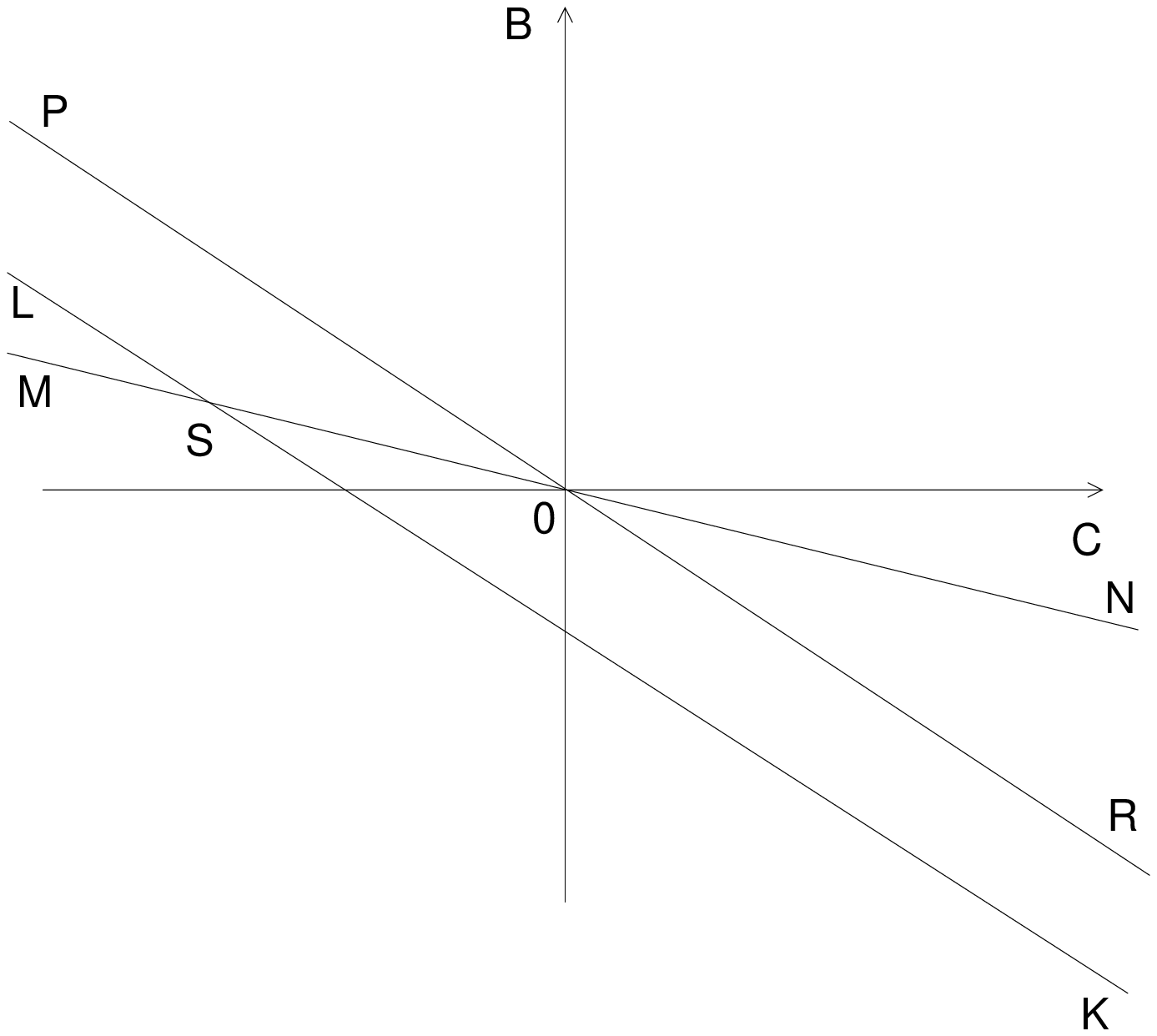}

   \label{f1}
\caption{Zones of de Sitter stability for the action (9). The lines $LK$ and
$PR$ are the same as in Fig.1; the line $MN$ is the $E=0$ line. The regular
de Sitter (if exists) is stable above this line, the singular de Sitter is stable
below this line}
  \end{figure}

\section{Bianchi I Universe in $3+1$ dimensions}

The situation become less simple even in $3+1$ dimensions
if we allow anisotropic perturbations of the de Sitter
solution, which means that the Weil tensor has
a non-zero contribution to the equations of motion 
\footnote{Main results of this section have been
found independently using another set of variables
in \cite{Hervik}.}.

Consider a flat homogeneous anisotropic Universe with the
Bianchi I metrics in the form

\begin{equation}
\mathrm{g}_{ik}=diag(-n(t)^2,a(t)^2,b(t)^2,c(t)^2).
 \label{8}
\end{equation}

It should be noted that in the presence of general quadratic curvature
corrections the diagonal form of the metrics (\ref{8}) is not the general one
\cite{Schmidt2}. We restrict ourself by this special form for simplicity. We show
that even in this less general case we have additional restrictions for stability
conditions of de Sitter solution.

Starting from the action (3)
we have three equations of motion
\begin{equation}
\begin{array}{l}
2\dot H_b+2\dot H_c +2H_b^2+2H_c^2+2H_cH_b -L +2B(H_b^{(3)}+3\dot
H_b^2 +2H_a^{(3)}+3\dot H_a^2 +H_a^2\dot H_b\\
+3\ddot H_cH_c+3\dot H_c^2+H_c^{(3)}+H_b^4 -H_a^4 +5H_c\dot
H_aH_a+H_aH_c\dot H_b-H_b^2H_aH_c-H_c^2H_aH_b\\
+H_bH_a^2H_c
+4H_c\dot H_bH_b+H_c^4+4\dot H_cH_c^2 +4H_c\ddot
H_a+2H_c\ddot H_b+H_a\ddot H_b+2H_b\ddot H_c\\
-H_c^3H_a+H_c^3H_b-H_a^3H_c-H_a^3H_b +H_b^3H_c-H_b^3H_a
+2H_c^2H_b^2 +H_a\ddot H_c-2H_a^2\dot H_a\\
+4H_a\ddot H_a+4H_b^2\dot H_b+3H_b\ddot H_b+4H_b\ddot H_a+3\dot
H_a\dot H_b +\dot H_aH_b^2 +3\dot H_a\dot H_c+\dot
H_aH_c^2\\+H_a^2\dot H_c +3\dot H_b\dot H_c +2\dot
H_bH_c^2+2H_b^2\dot H_c-H_a\dot H_bH_b+4H_b\dot
H_cH_c-H_a\dot H_cH_c\\
+H_bH_a\dot H_c+5H_b\dot H_aH_a+4H_cH_b\dot H_a)
+4C(2H_c^{(3)}-H_c^2H_a^2+6\ddot H_cH_c+H_b^4\\+5\dot
H_c^2-H_a^4+H_c^4 +2H_a^{(3)}+3\dot H_a^2+2H_b^{(3)}+5\dot
H_b^2-2H_bH_a^2H_c+2H_c^3H_b -2H_a^3H_b\\
-2H_a^3H_c+2H_b^3H_c-H_b^2H_a^2+3H_c^2H_b^2+2H_a\ddot H_c
 +2H_a\ddot H_b+4H_c\ddot H_b
+4H_b\ddot
H_a\\
+6H_b^2\dot H_b+6H_b\ddot H_b -2H_a^2\dot H_a+4H_a\ddot H_a
+4H_c\ddot H_a+6\dot H_cH_c^2+4\dot H_a\dot H_b +2\dot
H_aH_b^2\\
+6\dot H_b\dot H_c+4\dot H_bH_c^2 +4H_b^2\dot H_c+4H_b\ddot H_c
+4\dot H_a\dot H_c +2H_a\dot H_cH_c +2H_a\dot H_bH_b\\+2H_bH_a\dot
H_a +4H_cH_b\dot H_a+2\dot H_aH_c^2 +8H_c\dot H_bH_b+2H_c\dot
H_aH_a+2H_bH_a\dot H_c \\+8H_cH_b\dot H_c +2H_cH_a\dot H_b) = 0,
\end{array}
 \label{8.1}
\end{equation}
where $H_a=\dot a/a$, $H_b=\dot b/b$, $H_c=\dot c/c$,
two others can be obtained by transmutation $a\leftrightarrows c$ and
$a\leftrightarrows b$.

Substituting the de Sitter solution ($H_a=H_b=H_c=H_0=$Const),
we get that
the de Sitter solution has its unperturbed form $H_0=\sqrt{\Lambda/6}$.
After evaluating the corresponding system of equation of motion,
we find the following eigenvalues

\begin{equation}
\mu_{1,2,3}=-3H_0,
 \label{9}
\end{equation}
\begin{equation}
\mu_{4,5}=\frac{-18CH_0-6BH_0\pm2\sqrt{81C^2H_0^2+54BCH_0^2+9B^2H_0^2-6C-2B}}{2(6C+2B)},
 \label{10}
\end{equation}
\begin{equation}
\mu_{6,7,8,9}=\frac{-3BH_0\pm\sqrt{33B^2H_0^2+4B+96BCH_0^2}}{2B}.
 \label{11}
\end{equation}

In expanding Universe we have always $\mu_{1,2,3}<0$.
The condition  
$\mu_{4,5}<0$ leads to the already known result $B+3C>0$.
However, we have now additional restrictions arising
from the condition $\mu_{6,7,8,9}<0$.
It leads to
$\Lambda B+4C \Lambda +1<0$ if $B>0$ and $\Lambda |B|-4C \Lambda
-1<0$ if $B<0$.
Note that these stability conditions now depend not only on coefficients
before the higher-order terms in
the action (3), but also on the value of the cosmological constant $\Lambda$.
These additional restrictions arise from instability with respect to anisotropic
perturbations of the de Sitter metrics.  
It is interesting that independently of $\Lambda$ the whole quadrant
$B>0$, $C>0$, which always satisfies the stability conditions
in a FRW Universe, is excluded in anisotropic case (see Fig.3)

\begin{figure}[h]

\includegraphics[width=1\textwidth]{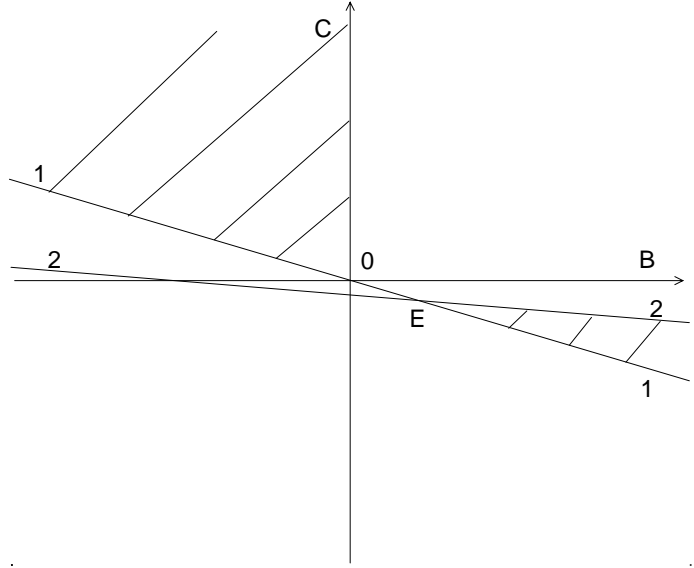}

   \label{f1}
\caption{Zones of de Sitter stability for the action (3) (shaded)
The de Sitter solutions in zone below the line $1$ are unstable
on isotropic background, the solutions in the unstable zone above the lines $1$ and $2$
are unstable due to anisotropic perturbations.  }
  \end{figure}

In the representation (8) we have the following conditions
$4 \tilde C \Lambda < \tilde B \Lambda/6-1$ for $\tilde B>0$
and $4 \tilde C \Lambda > |\tilde B|\Lambda/6-1$ for $\tilde B<0$.
(see Fig.4).

\begin{figure}[h]
\includegraphics[width=1\textwidth]{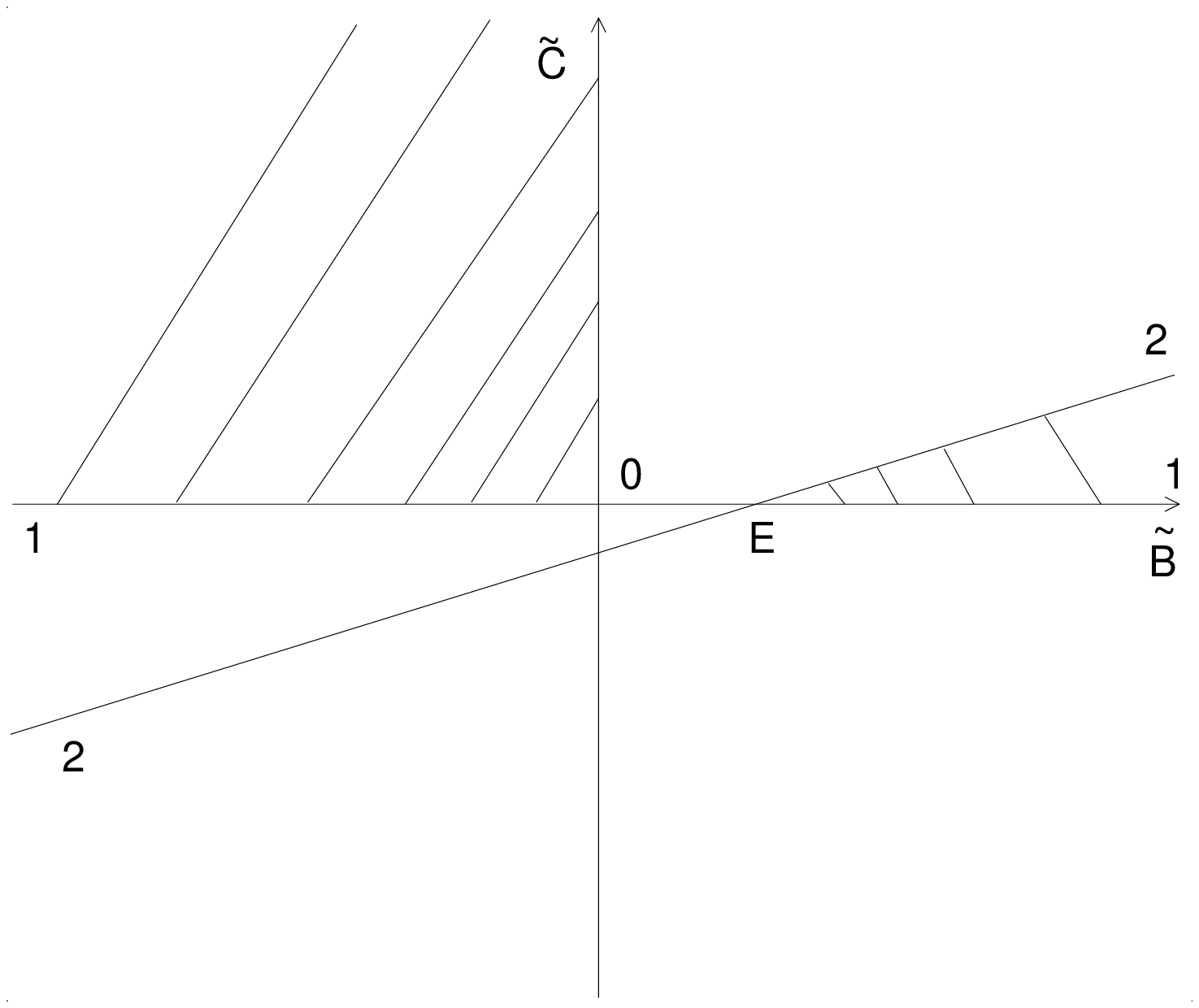}

   \label{f2}
\caption{Zones of de Sitter stability for the action (8) (shaded)
The unstable zone in the upper half-plane appears due to anisotropic
perturbations.}
  \end{figure}

\section{Conclusions}
We have studied the influence of second order curvature corrections in their general
form on stability of de Sitter solution. Our results for the $3+1$ dimensional space-time
and for higher-dimensional  cases differs significantly. The reason 
of this difference is non-zero contribution from the Gauss-Bonnet
combination for higher-dimensional space-times. In $3+1$ dimension the de Sitter
solution always exists, though it may be stable or unstable depending on particular form
of the second order corrections. The stability condition takes a very simple 
and independent of the value of cosmological constant $\Lambda$ form for
isotropic Universe, though become more complicated in anisotropic Universe.

In multidimensional Universe even in isotropic case (the only considered in the present
paper) there are several possibilities. Depending on the action we may have two de Sitter
solutions ( in this case one solution is stable and one solution is unstable), one de Sitter
solution (which can be either stable or unstable), and there are theories with no de Sitter
solutions at all.  

There results can be useful in constructions of cosmological scenarios taking into account
higher-derivative terms. Instability of a regular de Sitter solutions can damage the standard
inflationary scenario when close to de Sitter inflationary stage is produced by an effective
cosmological constant originated in a matter sector of the theory. This
opens a principal possibility to
rule out some theories of this type from the observational point of view. 
 On the other hand, unstable
singular de Sitter solution can be used in 
alternative approach to inflation \cite{Starobinsky}.
Our results show that quadratic curvature corrections in 
Lagrangian can produce such kind of  solutions
only in multidimensional Universe.

\section*{Acknowledgments}

This work is supported by RFBR grant 05-02-17450
and scientific school grant 2338.2003.2 of the Russian Ministry
of Science and Technology. Autors are grateful to Alexey
Starobinsky, Alan Coley, Sante Carloni, Sigbjorn Hervik and Hans-J\"urgen
Schmidt for discussions.

\end{document}